\documentclass[10pt]{article}

\usepackage{epsfig,graphicx}
\usepackage{color,colordvi}
\usepackage{amsmath,amssymb,mathabx}
\usepackage{lscape,slashed,rotating}
\usepackage{longtable,multirow,multicol}
\usepackage[applemac]{inputenc} 
\usepackage{subfigure}
\usepackage{adjustbox}
\usepackage{geometry} 
\begin{document}

\begin{center}

{\LARGE\bf Characterization and Optimization 

\vspace*{7pt}
of 

\vspace*{7pt}
Polarized and Unpolarized Positron Production}

\vspace*{30pt}

{\large\bf Sami Habet$^{ 1,2}$, Andriy Ushakov$^{ 1,2}$, Eric Voutier$^{ 1}$}

\vspace*{7pt}

{\it $^{1 }$Universit\'e Paris-Saclay, CNRS/IN2P3, IJCLab \\
15 rue Georges Cl\'emenceau, 91405 Orsay, France}

\vspace*{5pt}

{\it $^{2 }$Thomas Jefferson National Accelerator Facility \\
12000 Jefferson avenue, Newport News, VA 23606, USA}

\vspace*{10pt}

{{\bf PEPPo TN-23-01}, {\bf JLAB-ACC-23-3794}}

\vspace*{20pt}

\end{center}

% ----------------------------------------------------------------------------------------

\hrule

\begin{center}

{\large Abstract}

\end{center}

{\it The electron induced production of positrons is a multi-parameter problem which combines elementary physics  processes with complex collection systems. The optimization of  this technique for 120~MeV and 1000~MeV electron beam kinetic energies is here discussed considering a tungsten target. A strong correlation between the optimum target thickness and the angular acceptance ($\Delta \theta$) of the  collection system is observed, as well as sizeable differences between the optimum unpolarized and polarized operational modes. These also extend to the positron momentum and the positron polarization at the optimum, as well as the benefit of high electron beam energies for a high duty cycle positron source.}

\vspace*{10pt}

\hrule

\vspace*{10pt}

% ----------------------------------------------------------------------------------------

\section{Introduction}

The generation of positrons from the interaction of an electron beam in a high $Z$ target is the method selected for the production of positron beams at CEBAF~\cite{Hab22}, especially because it allows an efficient transfer of the polarization of the initial electron beam to the secondary positron  beam~\cite{Abb16} and is particularly suited to high-duty cycle beams. This technique proceeds through two distinct processes which occur in a single or two separate targets: firstly the radiation of photons via the bremsstrahlung of the electron beam, and secondly the creation of $e^+e^-$-pairs from the radiated photons. 

The present study reports about the optimization of this two step production for a single target scheme, in the perspective of the generation of polarized and unpolarized positron beams. The next sections revisit the elementary polarized bremsstrahlung and pair creation processes, particularly characterizing their angular distributions. The following section discussed the electron induced pair creation, specifically the effects of thick targets and the parameters characterizing the positron production. The optimization procedure is then discussed and applied for 120~MeV and 1000~MeV electron beam kinetic energies. The last section discusses the sensitivity of this production method to the electron beam energy.   

% ----------------------------------------------------------------------------------------

\section{Bremsstrahlung production of polarized photons}

As an elementary process of Quantum Electro-Dynamics (QED), the radiation of photons by electrons inside the environment of the nuclear field of a nucleus has been thoroughly studied over decades of theoretical and experimental investigations (see Ref.~\cite{Hau04} for a comprehensive  review). Hereafter, the production of unpolarized and polarized photons through the bremsstrahlung of unpolarized and polarized electrons or positrons is considered within the framework of the ultrarelativistic approach of Y.S.~Tsai~\cite{Tsa74}, consistent with the earlier calculations of H.~Olsen and L.C.~Maximon~\cite{Ols59} and the later finite lepton mass calculations of E.A.~Kuraev {\it et al.}~\cite{Kur10}.

\subsection{Cross section}

\begin{figure}[h!]
    \centering
    \includegraphics[width=0.95\columnwidth]{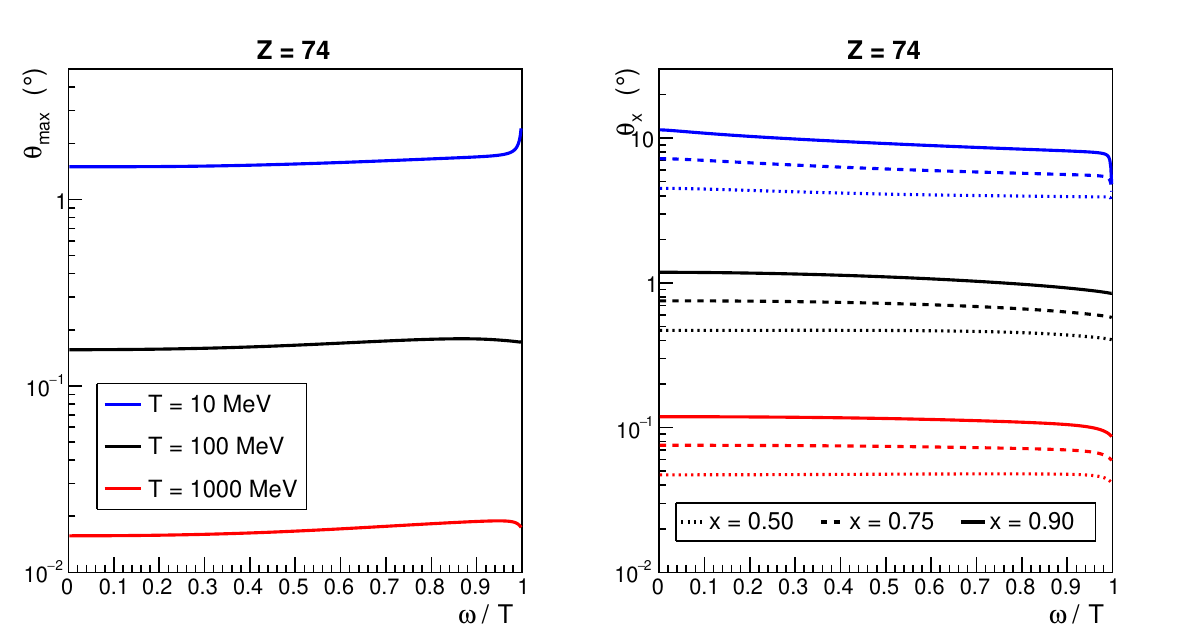}
    \caption{Characterization of the angular distribution of the 2-fold bremsstrahlung differential cross section for different electron beam energies: photon energy fraction dependency of the angle corresponding to the maximum cross section (left) and of the optimum angle corresponding to the 1-$x$ fraction of the maximum cross section (right).}
    \label{fig:angb}
\end{figure}
The energy ($\omega,k$) angle ($\Omega_k \equiv (\theta_k,\phi_k)$) distribution of the bremsstrahlung radiation emitted by a lepton beam of total energy $E$ in the field of a nucleus of electric charge $Z$ can be expressed as~\cite{Tsa74}
\begin{equation}
\frac{d \sigma_b}{dk \, d\Omega_k} =  A_0 \,
 \left[ \frac{2y-2}{{(1+l)}^2} + \frac{12l(1-y)}{{(1+l)}^4} \right] + 
 A_1 \, \left[ \frac{2 - 2y + y^2}{{(1+l)}^2} - \frac{4l(1-y)}{{(1+l)}^4} \right] \label{eqbrem}
\end{equation}
where $l$=$(E \sin(\theta_k) / m_e )^2$ denotes the angular dependency of the distribution with $m_e$ the electron mass, and $y$=$\omega/E$ is the photon energy fraction. In this expression
\begin{eqnarray}
A_0 & = & \frac{2 \alpha r_0^2}{\pi k} \, \frac{E^2}{m_e^2} \, (Z^2 + Z) \label{eqA0} \\
A_1 & = & \frac{2 \alpha r_0^2}{\pi k} \, \frac{E^2}{m_e^2} \, (X - 2 Z^2f) \label{eqA1}
\end{eqnarray}
are angle independent quantities related to the nuclear ($Z^2$ term in $A_0$) and electronic ($Z$ term in $A_0$) contributions to the bremsstrahlung cross section as well as to screening effects ($A_1$ term), as defined in Ref.~\cite{Tsa74}. The angular distribution 
\begin{equation}
\frac{d \sigma_b}{dk \, d\theta_k} = 2 \pi \sin(\theta_k) \, \frac{d \sigma_b}{dk \, d\Omega_k}
\end{equation}
is maximum at a $\theta_{max}$ angle roughly independent of the photon momentum (see Fig.~\ref{fig:angb} left) and which average is fairly approximated by the expression
\begin{equation}
\theta_{max} \approx \frac{1}{2} \, \frac{m_e}{E} \, .
\end{equation}
The $x$ optimum angle can then be defined as the angle  $\theta_{x}$ such that the energy angle differential cross section at this angle is the fraction 1-$x$ of the one at $\theta_{max}$. The photon momentum dependency of $\theta_{x}$ is shown on the right panel of Fig.~\ref{fig:angb} for typical electron beam kinetic energies in the range 10-1000~MeV. $\theta_{x}$ appears to be roughly independent of the photon momentum and, as expected, is strongly sensitive to $x$: at 100~MeV, a 1.5$^{\circ}$ cone contains 90\% of the produced photons while a 0.5$^{\circ}$ cone contains only 50\%.

\subsection{Polarization transfer}

\begin{figure}[b!]
    \centering
    \includegraphics[width=0.95\columnwidth]{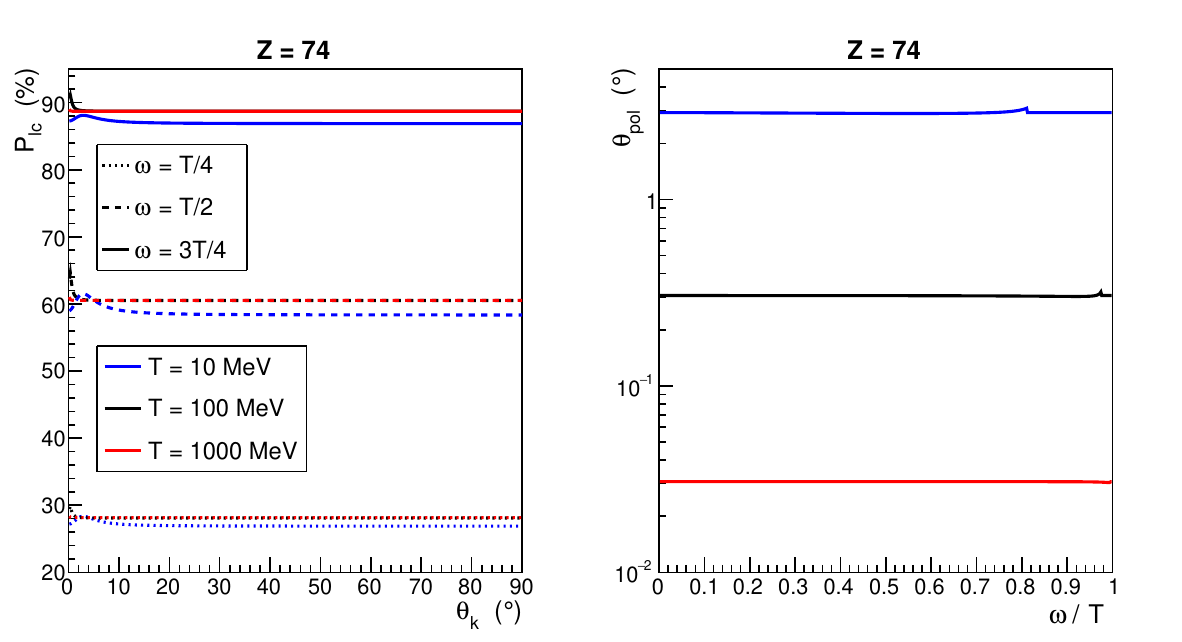}
    \caption{Angular distribution of the longitudinal-to-circular polarization transfer at different photon energy fractions (left), and energy dependence of the angle corresponding to the maximum polarization transfer (right), for different electron beam kinetic energies. The small discontinuity observed in the upper part of the kinematic region in the right panel is a  consequence of the limitations of the ultra-relativistic approximation~\cite{Dum09}.}
    \label{fig:bpol}
\end{figure}
Polarization effects in the bremsstrahlung process are  characterized by the self linear polarization of the  photon radiation emitted by an unpolarized electron beam, and by the circular polarization of the photon emitted by an initial electron beam polarized in the reaction plane (longitudinally or perpendicularly to the beam direction)~\cite{Ols59}. Following the dominance of the small angle region in the differential cross section, early calculations~\cite{Ols59} of these effects have been worked-out within the small angle approximation combined with an ultra-relativistic approach of the electron kinematics. The validity of the latter approximation was proved to be limited to a restricted region of the physics phase space~\cite{Dum09}, which was later superseded by finite electron mass calculations however within the peripheral kinematics limit~\cite{Kur10}. Full expressions  free of any approximation have been derived and will be soon available~\cite{Bys22}. Within the ultra-relativistic approach of H.~Olsen and L.C.~Maximon~\cite{Ols59}, the transfer $P_{lc}$ of the longitudinal polarization of an electron beam into circular polarization of bremsstrahlung photons can be expressed as 
\begin{equation}
P_{lc} = y \, \left[ 1 +  (X - 2 Z^2f) \frac{y(1-y)}{I_b}  \right]
\end{equation}
where
\begin{equation}
 I_b = Z^2 \, \left[ 2y - 2 + \frac{12l(1-y)}{{(1+l)}^2} \right] + 
 (X - 2 Z^2f) \, \left[ 2 - 2y + y^2 - \frac{4l(1-y)}{{(1+l)}^2} \right]
 \end{equation}
is proportional to the nuclear part of the unpolarized cross section Eq.~\eqref{eqbrem}. Distinctly from the cross section, the longitudinal polarization transfer is almost insensitive to the emission angle of the radiation (Fig.~\ref{fig:bpol} left), though there exists an angle $\theta_{pol}$ at which it is maximum. It is fairly approximated by the expression 
\begin{equation}
\theta_{pol} \approx \frac{m_e}{E} \, ,
\end{equation}
independently of the photon energy (Fig.~\ref{fig:bpol} right). The energy dependence of the polarization transfer should be noted: while it is strongly depending on the photon energy fraction, it appears roughly independent of the energy of the initial electron beam and leads to a universal S-shape behaviour in the photon energy fraction variable~\cite{Ols59}.

% ----------------------------------------------------------------------------------------

\section{Creation of polarized letpon pairs}

As a reciprocal process, the pair creation reaction was  investigated along with the bremsstrahlung taking advantage of kinematical substitution properties. This feature is most likely the origin of the unphysical behaviour of 
polarization transfers~\cite{Dum09} of earlier calculations~\cite{Ols59}, because transporting  inappropriately the ultra-relativistic approximation. However, the unpolarized cross section does not suffer these limitations and was consistently derived from either first principles~\cite{Tsa74} or substitution properties~\cite{Hab21}.

\subsection{Cross section}

\begin{figure}[ht!]
    \centering
    \includegraphics[width=0.95\columnwidth]{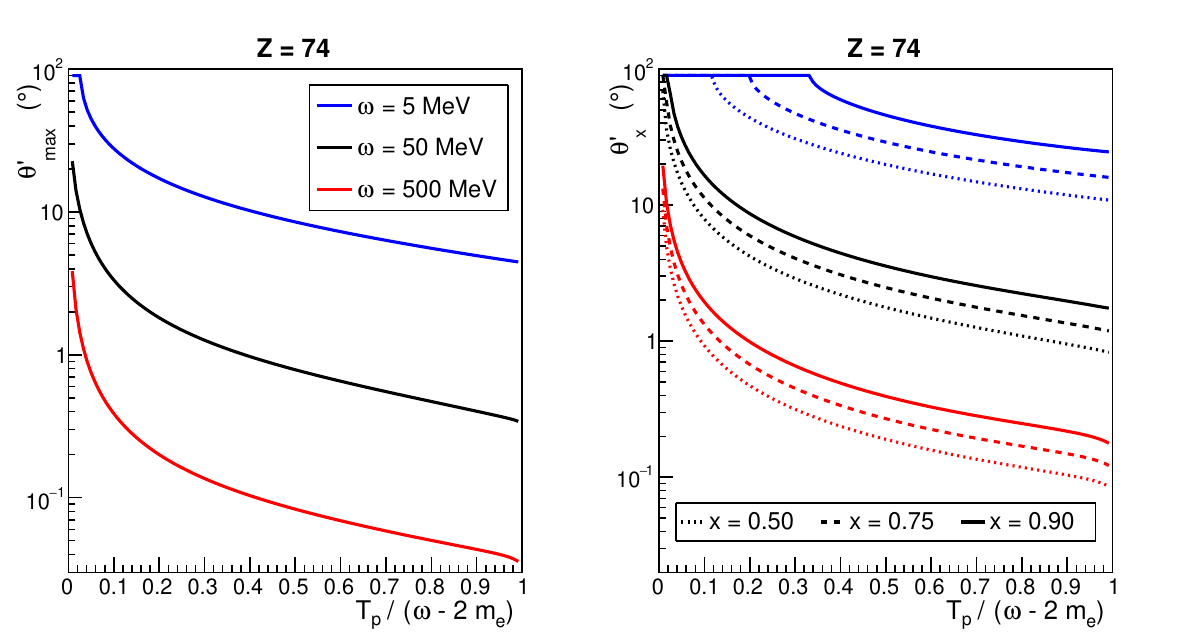}
    \caption{Characterization of the angular distribution of the 2-fold differential cross section of the pair creation process for different initial photon energies: positron kinetic energy fraction dependency of the angle corresponding to the maximum cross section (left) and of the optimum angle corresponding to the 1-$x$ fraction of the maximum cross section (right). The saturation at 90$^{\circ}$ indicates that the cross section at the corresponding positron energy is always larger than the selected $\theta_x$-cross section.}
    \label{ang_pair}
\end{figure}
The energy ($E,p$) angle ($\Omega_p \equiv (\theta_p,\phi_p)$) distribution of the positrons emitted by a photon beam of energy $\omega$ in the field of a nucleus of electric charge $Z$ can be expressed as~\cite{Tsa74}
\begin{equation}
\frac{d \sigma_p}{dp \, d\Omega_p} =  B_0 \,
 \left[ \frac{2y'(1-y')}{{(1+l)}^2} - \frac{12ly'(1-y')}{{(1+l)}^4} \right] + 
 B_1 \, \left[ \frac{1 - 2y' + 2y'^2}{{(1+l)}^2} + \frac{4ly'(1-y')}{{(1+l)}^4} \right] \label{eqpair}
\end{equation}
where $l$=$(E \sin(\theta_p) / m_e )^2$ denotes the angular dependency, and $y'$=$E/\omega$ is the positron energy fraction. In this expression
\begin{eqnarray}
B_0 & = & \frac{2 \alpha r_0^2}{\pi \omega} \, \frac{E^2}{m_e^2} \, (Z^2 + Z) \\
B_1 & = & \frac{2 \alpha r_0^2}{\pi \omega} \, \frac{E^2}{m_e^2} \, (X - 2 Z^2f)
\end{eqnarray}
are angle independent quantities related to the nuclear ($Z^2$ term in $B_0$) and electronic ($Z$ term in $B_0$) contributions to the pair creation cross section as well as to screening effects ($B_1$ term), similarly to $(A_0,A_1)$ defined in Eq.~\eqref{eqA0} and Eq.~\eqref{eqA1}. 
As for the bremsstrahlung case, the angular distribution 
\begin{equation}
\frac{d \sigma_p}{dp \, d\theta_p} = 2 \pi \sin(\theta_p) \, \frac{d \sigma_p}{dp \, d\Omega_p}
\end{equation}
of the produced positrons exhibits a maximum at $\theta'_{max}$, however strongly depending on the positron energy (Fig.~\ref{ang_pair} left). The $x$ optimum angle $\theta'_x$ can also be defined as the angle corresponding to the fraction 1-$x$ of the cross section at $\theta'_{max}$. The sensitivity of this angle to the positron energy fraction (Fig.~\ref{ang_pair} right) strongly differs from the bremsstrahlung case. It can empirically be parameterized by the expression 
\begin{equation}
\theta'_x (\omega,T_p) \equiv a_0 \, \exp{ \left[ 
\frac{a_1}{\frac{T_p}{\omega-2m_e}-a_2}  
+ \frac{a_3 T_p}{\omega-2m_e} 
+ a_4 \left( \frac{T_p}{\omega-2m_e} \right)^2
+ a_5 \left( \frac{T_p}{\omega-2m_e} \right)^3
\right] } 
\end{equation}
where $a_i \equiv a_i(\omega,x)$. At small photon and positron energies, $\theta'_{x}$ is particularly large  corresponding to a 4$\pi$ efficient emission of positrons. As the photon energy increases, the Lorentz boost of final state particles impacts the angular distribution of positrons and $\theta'_{x}$ is reduced. The decrease of $\theta'_{x}$ with the positron energy fraction is a consequence of the limited angular space for high energy positrons. These are typical kinematic  characteristics of the decay of a system into 2 particles with identical mass. 

\subsection{Polarization transfer}

\begin{figure}[h!]
    \centering
    \includegraphics[width=0.95\columnwidth]{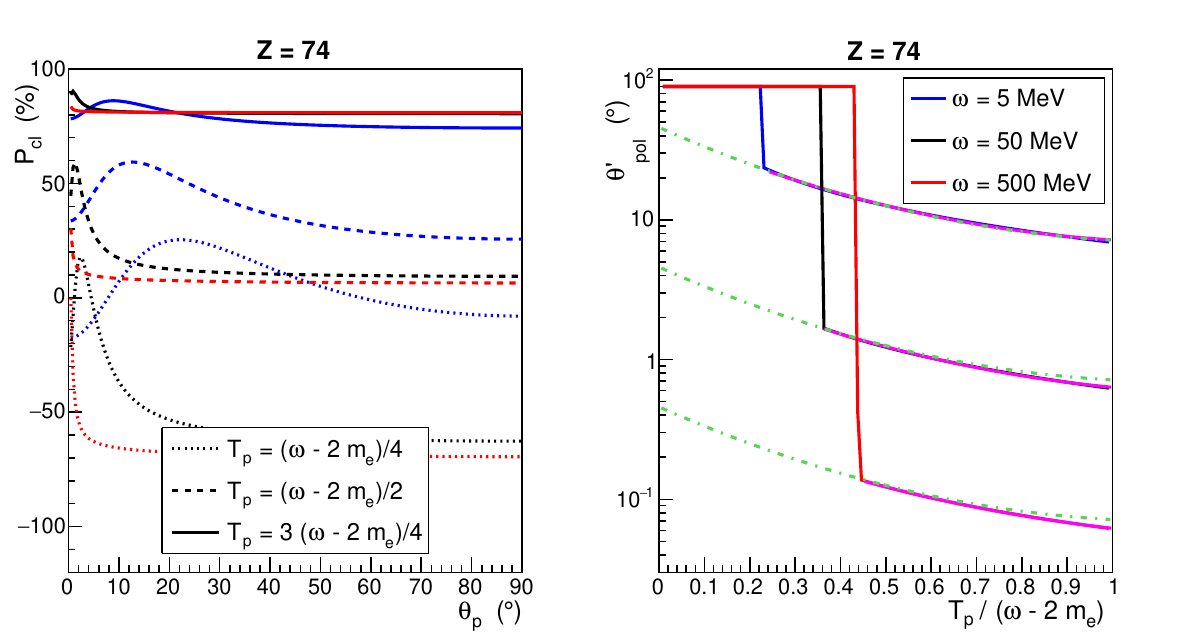}
    \caption{Angular distribution of the circular-to-longitudinal polarization transfer at different positron energy fractions (left), and energy dependence of the angle corresponding to the maximum polarization transfer (right), for different photon energies.}
    \label{pol_pair}
\end{figure}
Polarization effects in the pair creation process are uniquely  characterized by the transfer of the circular polarization of  photons into longitudinal and perpendicular polarization components of the pair. Similarly to bremsstrahlung, the  circular-to-longitudinal transfer is more efficient than the circular-to-perpendicular transfer. The early calculations of these effects within the ultra-relativistic and small angle  approximations~\cite{Ols59} have been shown to be severely limited~\cite{Dum09} and an empirical {\it ad hoc} solution was developped to correct for the observed unphysical behaviours~\cite{Dum11}. In this framework, the circular-to-perpendicular polarization transfer $P_{cl}$ can be expressed as~\cite{Ols59}
\begin{equation}
P_{cl} = \frac{1}{y'} \, \left[ 1 + (X - 2 Z^2f) \, \frac{1-y'}{I_p} \right]
\end{equation}
where
\begin{equation}
 I_p = Z^2 \, \left[ 2y'(1-y') - \frac{12ly'(1-y')}{{(1+l)}^2} \right] +  (X - 2 Z^2f) \, \left[ 1 - 2y' + 2y'^2 + \frac{4ly'(1-y')}{{(1+l)}^2} \right] \, .
\end{equation}
The angular distribution of $P_{cl}$ (Fig.~\ref{pol_pair}) is strongly sensitive to the positron energy fraction while the photon energy affects the magnitude of the polarization transfer in a significant manner as the positron energy decreases. The angle $\theta'_{pol}$ at which the polarization transfer is maximum is also strongly depending on the positron energy fraction. Large $\theta'_{pol}$ angles appear relevant of the low energy part of the positron spectra, consistently with a weak angular dependence of $P_{cl}$. The absence of a smooth transition between the low and high energy part of positron spectra reflects the limitations of the ultra-relativistic approximation. Full calculations free of any approximation will be soon available~\cite{Bys22}. In the high energy region, the $\theta'_{pol}$ sensitivity to the photon and positron energies is well parameterized by the expression
\begin{equation}
\theta'_{pol} (\omega, T_p) = b_0 \, \frac{m_e}{\omega} \, \exp{ \left[ \left( b_1  + b_2  \frac{T_p}{\omega-2m_e} \right) \frac{T_p}{\omega-2m_e} \right] }
\end{equation}
where the $b_i$ parameters for a tungsten nuclei are 
\begin{eqnarray}
b_0 & = & \phantom{-}44.9 \\
b_1 & = &           -3.32 \\
b_2 & = & \phantom{-}1.46 \, .
\end{eqnarray}
It should be noted that the high energy part of the positron spectra, which features the largest polarization transfer,  also features the smallest $\theta'_{pol}$ angles. Together with the angular dependence of the cross section it suggests, at the level of a single interaction process, the importance of the forward region for the production of polarized positron beams. 

% ----------------------------------------------------------------------------------------

\section{Electron induced pair production}

\subsection{Thick target effects} 

\begin{figure}[h!]
    \centering
    \includegraphics[width=0.50\columnwidth]{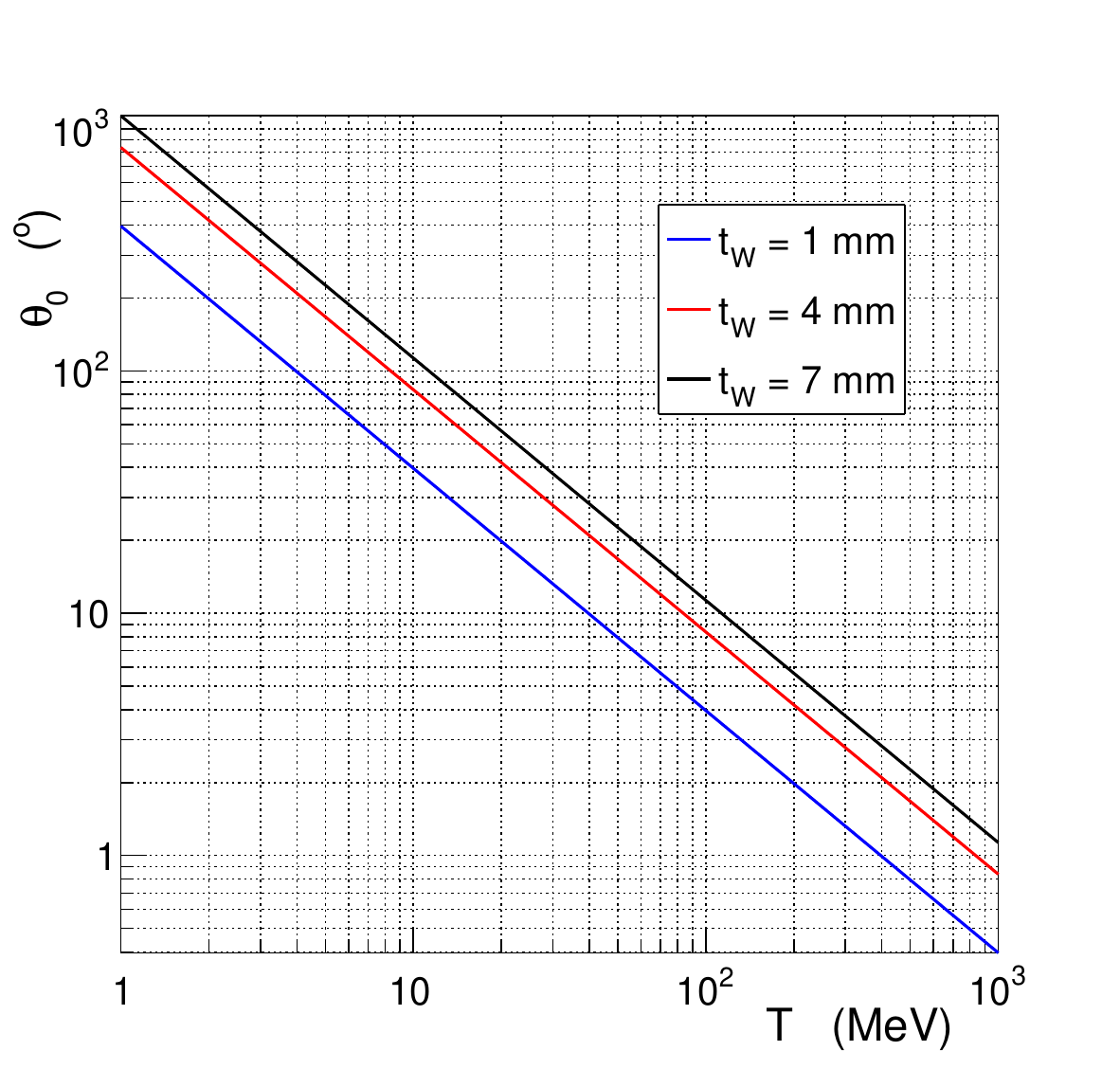}
    \caption{Energy dependence of the multiple scattering spread of an electron or positron distribution, for different thicknesses of tungsten material.}
    \label{fig:mulsca}
\end{figure}
The production of positrons from primary electrons combines the bremsstrahlung and pair creation processes within a single or a multiple target system. In that respect, the thicknesses of the targets become optimizable parameters with respect to the performance of the production scheme. The angular properties described in the previous sections are specific of thin targets (a few percent of radiation length) where the secondary particles do not suffer much distortion from the creation medium. In the case of thick targets, these features are completely swamped by multiple scattering effects which then govern most of the angular distribution of secondary particles and generate significant transverse momentum. The energy dependence of the average opening angle of an electron/positron distribution under influence of multiple scattering is represented on Fig.~\ref{fig:mulsca} for different thickness of tungsten material. This readily tells that multiple scattering effects distribute low energy particles over a very large angular domain while high energy particles remain in a reasonable down to minimal angular range.

\subsection{Characteristic parameters}

\begin{figure}[b!]
    \centering
    \includegraphics[width=0.80\columnwidth]{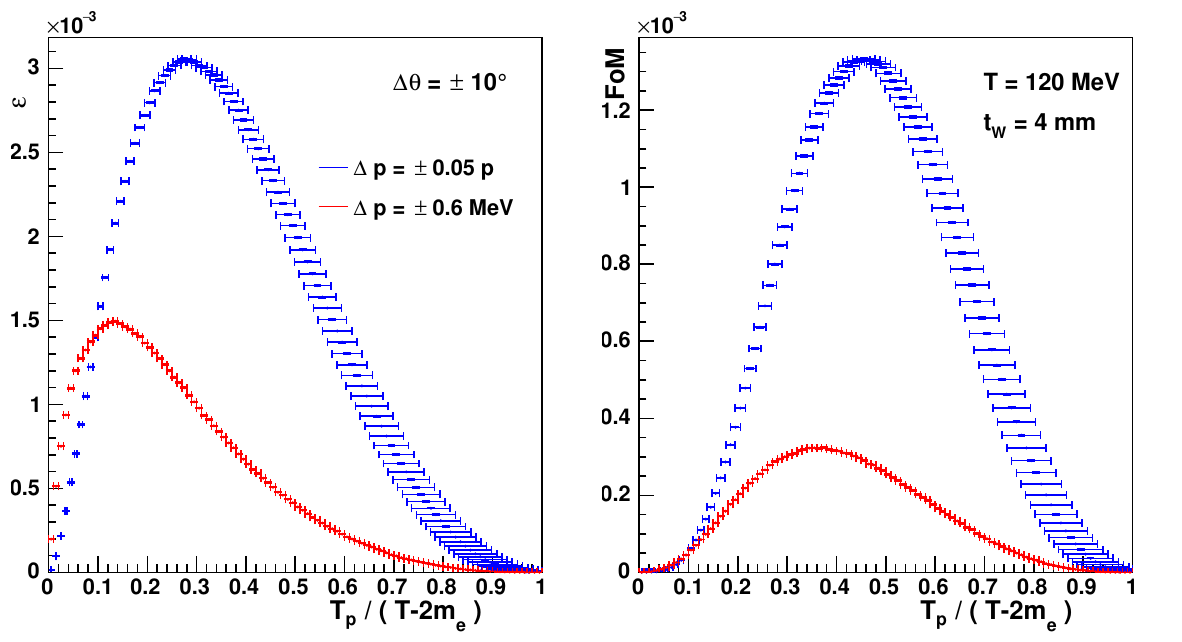}
    \caption{Energy dependence of the efficiency (left) and the  Figure-of-Merit (right) of the positron population produced by a 120~MeV polarized electron beam interacting with a 4~mm tick tungsten target. The different color  curves show the sensitivity of the shape of the distributions to the energy binning as defined in Eq.~(\ref{efeq})-(\ref{foeq}): constant (red) and variable (blue) binning.}
    \label{momdis}
\end{figure}
Considering a single target production scheme, the performance of an unpolarized secondary positron source is characterized by the production efficiency ($\varepsilon \equiv \varepsilon (p,\theta,\phi)$), which quantifies the positron yield ($N_p(p,\theta,\phi)$) at the momentum $p$ and spherical angles $(\theta,\phi)$ in comparison with the incident electron beam ($N_e$), that is
\begin{equation}
\varepsilon (p,\theta,\phi) = \frac{N_p(p,\theta,\phi)}{N_e} \, .
\end{equation}
The positron yield is determined as 
\begin{equation}
N_p(p,\Delta p,\Delta \theta) = N_e \int_{p-\Delta p}^{p+\Delta p} d\xi  \int_{0}^{\Delta \theta} d \theta \int_{0}^{2\pi} d \phi \,  \frac{d\varepsilon(\xi,\theta,\phi)}{d\xi \, d\theta \, d\phi}
\label{efeq}
\end{equation}
where $\Delta p$ represents an effective fixed ($\pm0.6$~MeV) or variable ($\pm 0.05\, p$) momentum acceptance, and similarly for $\Delta \theta$. The performance of a polarized source is better characterized by the Figure-of-Merit ($FoM \equiv FoM(p)$), which combines the efficiency and the averaged polarization ($P_p$) of particles according to the expression
\begin{equation}
FoM(p,\Delta p,\Delta \theta) = \frac{N_p(p,\Delta p,\Delta \theta)}{N_e} \, \langle P_p^2 \rangle =  \int_{p-\Delta p}^{p_+\Delta p} d \xi  \int_{0}^{\Delta \theta} d \theta \int_{0}^{2\pi} d \phi \,  \frac{d\varepsilon(p,\theta,\phi)}{d\xi \, d\theta \, d\phi} \, P_p^2(p,\theta,\phi) \, .
\label{foeq}
\end{equation}

The combination of the bremsstrahlung and pair creation processes produces a positron flux dominated by low energy particles. Since very low energy positrons do not easily escape the target, the flux of secondary particles most likely goes through a maximum at a specific positron energy. The left panel of Fig.~\ref{momdis} shows the momentum distribution of the positron production efficiency from a 120~MeV polarized electron beam in a 4~mm thick tungsten target. The distribution is limited to positrons emitted within a 10$^{\circ}$ cone angle which mimics the angular acceptance of a collection system. The observed maximum value of the distribution defines the energy domain of the particles to be collected. The Figure-of-Merit (Fig.~\ref{momdis} right) follows the same trend however with a shift of the positron momentum at maximum due to the energy dependence of the positron polarization (Fig.~\ref{pol_pair}). Another shift can be observed which characterizes the effect of a variable bin width as compared to a constant one. In an attempt to mimic the acceptance effect of a magnetic collection system on the positron yield, a variable bin width is introduced which is linked to the positron momentum acceptance $\Delta p / p$ ($\pm5$\% in the present case) of the collection system. This considerably distorts the energy distribution as compared to the constant bin width picture and indicates that optimum efficiency and Figure-of-Merit are obtained at larger positron momenta.

% ----------------------------------------------------------------------------------------

\section{Optimization procedure}

The maximum values of $\varepsilon$ and $FoM$, as well as the positron energy at maximum depend on the target thickness and on the incident electron beam energy. It is the purpose of the optimization study to determine the thickness which maximizes $\varepsilon$ and $FoM$ at a given beam energy by extending a method previously proposed and used in  Ref.~\cite{Dum11}. The procedure is based on extensive GEANT4~\cite{Ago03} simulations where the positron yield and Figure-of-Merit are investigated for different momentum and angular acceptances of the collection system assuming a pencil beam of electrons. At fixed target thickness and beam energy, the quantities of interest for the unpolarized operational mode are attached to the maximum efficiency as:
\begin{itemize}
\item{$\varepsilon_{max}$, the maximum positron production efficiency;}
\item{$FoM_{\varepsilon}$, the Figure-of-Merit at $\varepsilon_{max}$;}
\item{$p_{\varepsilon}$, the positron momentum in MeV/$c$ units at $\varepsilon_{max}$;}
\item{$P_{\varepsilon}$, the longitudinal polarization of positrons at $\varepsilon_{max}$.}
\end{itemize}
The quantities of interest for the polarized operational mode are similarly attached to the maximum Figure-of Merit as:
\begin{itemize}
\item{$FoM_{max}$, the maximum Figure-of-Merit;}
\item{$\varepsilon_{FoM}$, the positron production efficiency at $FoM_{max}$;}
\item{$p_{FoM}$, the positron momentum in MeV/$c$ units at $FoM_{max}$;}
\item{$P_{FoM}$, the longitudinal polarization of positrons at $FoM_{max}$.}
\end{itemize}
These parameters are determined from the basic spectra of the energy distribution of the efficiency, the average longitudinal polarization, and the Figure-of-Merit. 

\begin{figure}[t!]
    \centering
    \includegraphics[width=0.95\columnwidth]{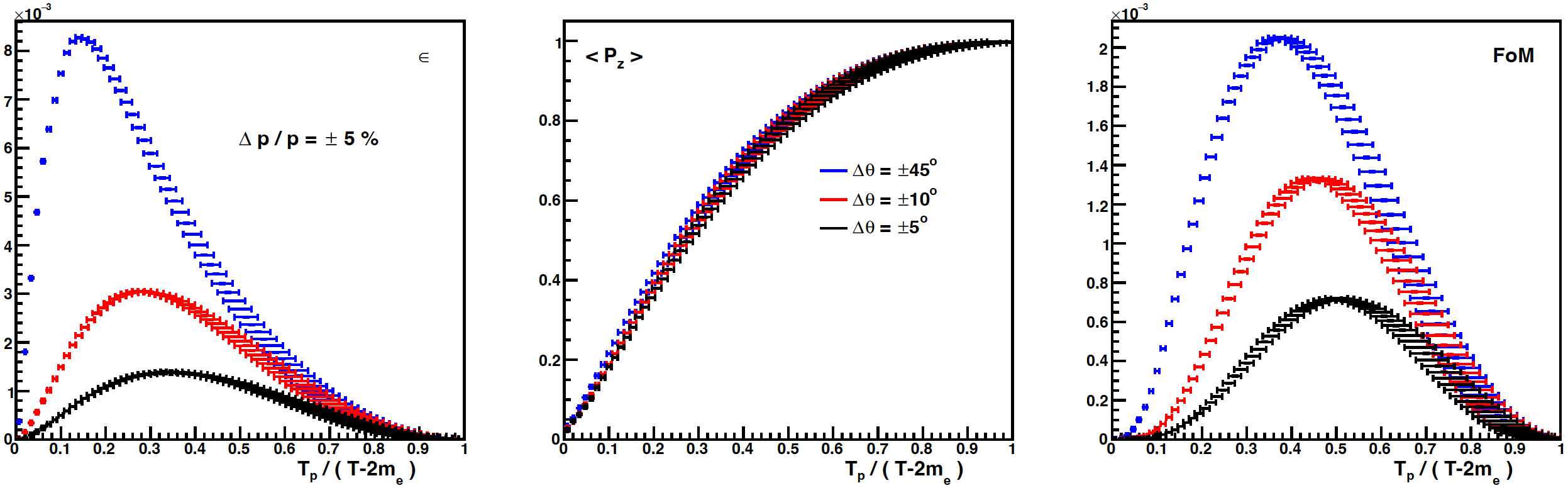}
    \caption{Positron energy distribution of the production efficiency (left panel), the average longitudinal polarization (middle panel), and the Figure-of-Merit (right panel), simulated for 120~MeV incident electrons onto a 4~mm thick tungsten target; different angular acceptances of the positron collection system are compared at a fixed $\pm5$\% positron momentum acceptance.}
    \label{opt_base}
\end{figure}

For example, the  parameters deduced from Fig.~\ref{opt_base} are reported in Tab.~\ref{opt:eff} and Tab.~\ref{opt:fom}. The error bars include both statistical and systematical effects in quadrature. While systematics on the maxima is determined as half of the bin width, that of positron momentum at maxima takes also into account the statistical fluctuation of the distribution which dominates for slowly varying quantities. The systematics on the longitudinal polarization is determined from that of the positron momentum at maxima following the energy/polarization correlation of Fig.~\ref{opt_base} (middle panel). From these tables, noticeable differences between the optimized unpolarized and polarized modes are observed which importance decreases with the angular acceptance. Because of multiple scattering effects, large angular openings  favor the acceptance of low energy positrons. The production efficiency is consequently maximum at low energy, that is in a region where the polarization is small. Following the energy dependence of the positron polarization, the Figure-of-Merit then maximizes at larger positron energy. Reducing the angular acceptance rejects more low than high energy particles which somehow favors the high energy part of the positron spectra. Therefore, the difference between unpolarized and polarized modes becomes less significant. The unpolarized operational mode appears consequently strongly correlated with the angular acceptance, although a non-zero polarization remains at the maximum efficiency.

\begin{table}[t!]
\centering
\begin{adjustbox}{max width=\textwidth}
\begin{tabular}{c||c|c|c|c||c|c|c|c|c|c}
$\Delta \theta$ & $\varepsilon_{max}$ & $p_{\varepsilon}$ & $P_{\varepsilon}$ & $FoM_{\varepsilon}$ & $\delta \varepsilon_{max}$ & $\delta p_{\varepsilon}^-$ & $\delta p_{\varepsilon}^+$ & $\delta P_{\varepsilon}^-$ & $\delta P_{\varepsilon}^+$ & $\delta FoM_{\varepsilon} $ \\
($^{\circ}$) & ($\times 10^{-3}$) & (MeV/$c$) & (\%) & ($\times 10^{-3}$) & ($\times 10^{-6}$) & (MeV/$c$) & (MeV/$c$) & (\%) & (\%) & ($\times 10^{-6}$) \\ \hline\hline
           45 & 8.27 & 17.2 & 29.4 & 0.72 & 8.91 & 1.80 & 1.80 & 2.58 & 2.56 & 1.20 \\ \hline
           10 & 3.05 & 33.0 & 50.9 & 0.79 & 5.40 & 2.29 & 3.58 & 2.18 & 4.23 & 1.66 \\ \hline
 \phantom{0}5 & 1.39 & 41.0 & 59.9 & 0.50 & 3.64 & 2.59 & 3.77 & 2.00 & 3.74 & 1.45 \\
\end{tabular}
\end{adjustbox}
\caption{Quantities of interest characterizing the unpolarized operational mode of a positron source based on a 120~MeV polarized electron beam  interacting within a 4~mm thick tungsten target, considering a constant positron momentum acceptance $\Delta p / p$=$\pm5$\% and different angular acceptances.}
\label{opt:eff}
\end{table}
\begin{table}[t!]
\centering
\begin{adjustbox}{max width=\textwidth}
\begin{tabular}{c||c|c|c|c||c|c|c|c|c|c}
$\Delta \theta$ & $FoM_{max}$ & $p_{FoM}$ & $P_{FoM}$ & $\varepsilon_{FoM}$ & $\delta FoM_{max}$ & $\delta p_{FoM}^-$ & $\delta p_{FoM}^+$ & $\delta P_{FoM}^-$ & $\delta P_{FoM}^+$ & $\delta \varepsilon_{FoM}$ \\
($^{\circ}$) & ($\times 10^{-3}$) & (MeV/$c$) & (\%) & ($\times 10^{-3}$) & ($\times 10^{-6}$) & (MeV/$c$) & (MeV/$c$) & (\%) & (\%) & ($\times 10^{-6}$) \\ \hline\hline
 45 & 2.05 & 44.1 & 66.2 & 4.68 & 3.16 & 2.72 & 3.87 & 1.76 & 3.33 & 6.69 \\ \hline
 10 & 1.33 & 55.2 & 76.2 & 2.30 & 2.81 & 4.21 & 4.21 & 2.96 & 2.76 & 4.68 \\ \hline
 \phantom{0}5 & 0.72 & 60.0 & 79.3 & 1.15 & 2.13 & 3.39 & 4.37 & 1.29 & 2.48 & 3.30 \\
\end{tabular}
\end{adjustbox}
\caption{Quantities of interest characterizing the polarized opertional mode of a positron source based on a 120~MeV polarized electron beam  interacting within a 4~mm thick tungsten target, considering a constant positron momentum acceptance $\Delta p / p$=$\pm5$\% and different angular acceptances.}
\label{opt:fom}
\end{table}

% ----------------------------------------------------------------------------------------

\section{Optimization of production at 120~MeV}

\subsection{Unpolarized operational mode}

\begin{figure}[h!]
\centering
\includegraphics[width=0.999\columnwidth]{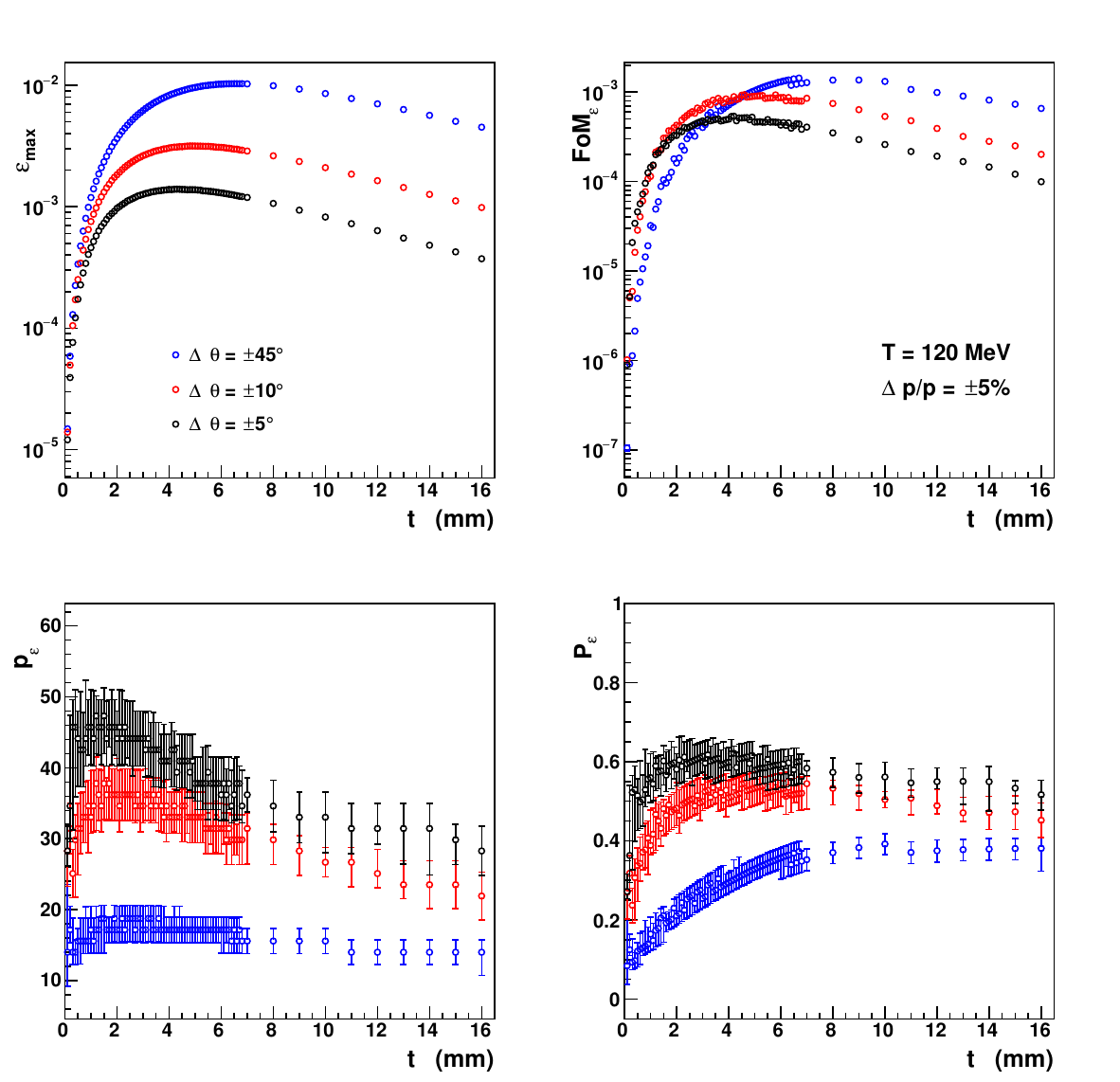}
\caption{Target thickness dependence of the characteristic  quantities of the unpolarized mode operation at 120~MeV and a momentum acceptance of $\pm$5\% for different angular acceptances.}
\label{opt120_unp}
\end{figure}

The sensitivity of the maximum positron production efficiency to the target thickness is here studied for a variable bin width corresponding to the positron momentum acceptance $\Delta p /p$=$\pm$5\% and different angular acceptances. The results ($\varepsilon_{max}, FoM_{\varepsilon}, p_{\varepsilon}, P_{\varepsilon}$) are reported on Fig.~\ref{opt120_unp} for target thicknesses varying up to 16~mm. The production efficiency is evidently small for thin targets and rapidly increases to reach a slowly varying optimum region. As the thickness increases, the production becomes less efficient because the attenuation of the positron flux is not compensated by the positron produced in the extra thickness. As dominated by the efficiency variation, the Figure-of-Merit follows a similar trend consistent with an almost constant positron polarization. It is only in the case of thin targets and large angular acceptances that an almost unpolarized positron flux is obtained. Large angular acceptances are hardly achieved in practice and therefore the optimized unpolarized mode features in fact a subsequently polarized flux.

The thickness $t_{\varepsilon}$ optimizing the positron production is given in Tab.~\ref{eff_max_120}. While $t_{\varepsilon}$ appears strongly sensitive to the angular acceptance, the dependence on the momentum acceptance is less pronounced. It is worth noted that the errors reported in the table results from a combination of the simulation statistics (1.05$\times10^8$ pencil electrons) and the shape of the distribution in the maximum region. These are determined from the comparison of the lower statistical edge of the maximum efficiency  ($\varepsilon_{max}(t_{\varepsilon}) -\delta\varepsilon_{max}(t_{\varepsilon})$) with the upper edge of other values at different thicknesses ($\varepsilon_{max}(t) +\delta\varepsilon_{max}(t)$). The flatness of the distribution around the optimum thickness leads to values larger than the step (0.1~mm) of the simulations. The uncertainty on the positron momentum at maximum also reflects this feature.

\begin{table}[h!]
\centering
\begin{adjustbox}{max width=\textwidth}
\begin{tabular}{c||c|c|c|c|c|c||c|c|c|c|c|c}
 \multirow{2}{*}{$\Delta \theta$} & \multicolumn{6}{c||}{$\Delta p / p = \pm 5 $\%} & \multicolumn{6}{c}{$\Delta p / p = \pm 10 $\%} \\
& $t_{\varepsilon}$ & $\delta t_{\varepsilon}^-$ & $\delta t_{\varepsilon}^+$ & $p_{\varepsilon}$ & $\delta p_{\varepsilon}^-$ & $\delta p_{\varepsilon}^+$ & $t_{\varepsilon}$ & $\delta t_{\varepsilon}^-$ & $\delta t_{\varepsilon}^+$ & $p_{\varepsilon}$ & $\delta p_{\varepsilon}^-$ & $\delta p_{FoM}^+$ \\
($^{\circ}$) & (mm) & (mm) & (mm) & (MeV/$c$) & (MeV/$c$) & (MeV/$c$) & (mm) & (mm) & (mm) & (MeV/$c$) & (MeV/$c$) & (MeV/$c$) \\ \hline\hline
          45 & 6.5 & 0.2 & 0.5 & 17.2 & 3.29 & 1.80 & 5.8 & 0.4 & 0.3 & 25.1 & 2.02 & 2.02 \\
          10 & 4.8 & 0.2 & 0.5 & 33.0 & 3.58 & 3.58 & 4.8 & 0.2 & 0.5 & 33.0 & 3.58 & 3.58 \\
\phantom{0}5 & 4.3 & 0.2 & 0.2 & 39.4 & 2.53 & 3.73 & 4.3 & 0.2 & 0.2 & 39.4 & 2.53 & 3.73 \\
\end{tabular}
\end{adjustbox}
\caption{Optimum target thickness $t_{\varepsilon}$ and positron momentum at maximum $p_{\varepsilon}$ of the unpolarized mode at 120~MeV, for different angular and momentum acceptances.}
\label{eff_max_120}
\end{table}

%---------------------------------------------------------------------------------------------

\subsection{Polarized operational mode}

\begin{figure}[h!]
\centering
\includegraphics[width=0.999\columnwidth]{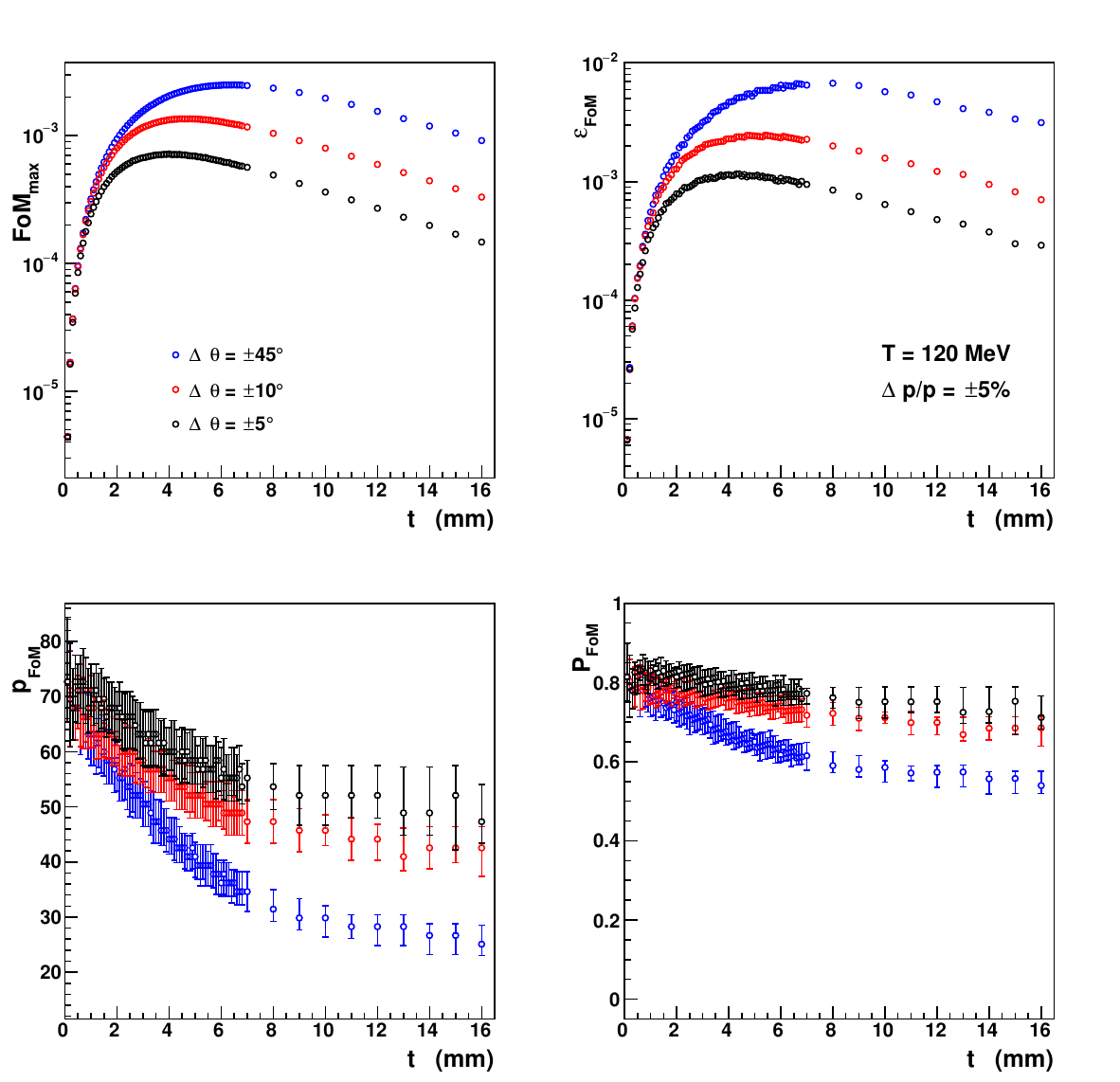}
\caption{Target thickness dependence of the characteristic  quantities of the polarized mode operation at 120~MeV and a momentum acceptance of $\pm$5\% for different angular acceptances.}
\label{opt120_pol}
\end{figure}

Under the same conditions than the unpolarized case, the dependence of the $FoM$ on the target thickness is  investigated up to 16~mm. The results ($FoM_{max}, \varepsilon_{FoM},p_{FoM}, P_{FoM}$) reported on Fig.~\ref{opt120_pol} are very similar to the unpolarized mode optimization except for thin targets where the positron momentum at maximum and consequently the positron beam polarization are initially large. For realistic angular acceptances the rule of thumb established at lower beam energies~\cite{Dum11} is still valid: the $FoM$ is maximum at about half of the electron beam energy, corresponding to a polarization transfer of about 75\%. It is remarkable to note that a high polarization level is obtained, almost independent of the target thickness. This was experimentally observed at the PEPPo experiment~\cite{Abb16}, and corresponds to multiple scattering effects which reduce the positron energy but do not affect their polarization. The positrons produced at $p_{FoM}$ in the first part of a thick target still loose energy when traveling through the second part and contribute to build-up an approximately constant average polarization. Finally, the  optimum target thickness is, similar to the unpolarized mode, roughly insensitive to the momentum acceptance but strongly depends on the angular one (Tab.~\ref{fom_max_120}). 

% ----------------------------------------------------------------------------------------

\newpage

\null\vfill
\vfill\eject

\null\vfill
\vfill\eject

\null\vfill
\begin{figure}[h!]
\centering
\includegraphics[width=0.76\columnwidth]{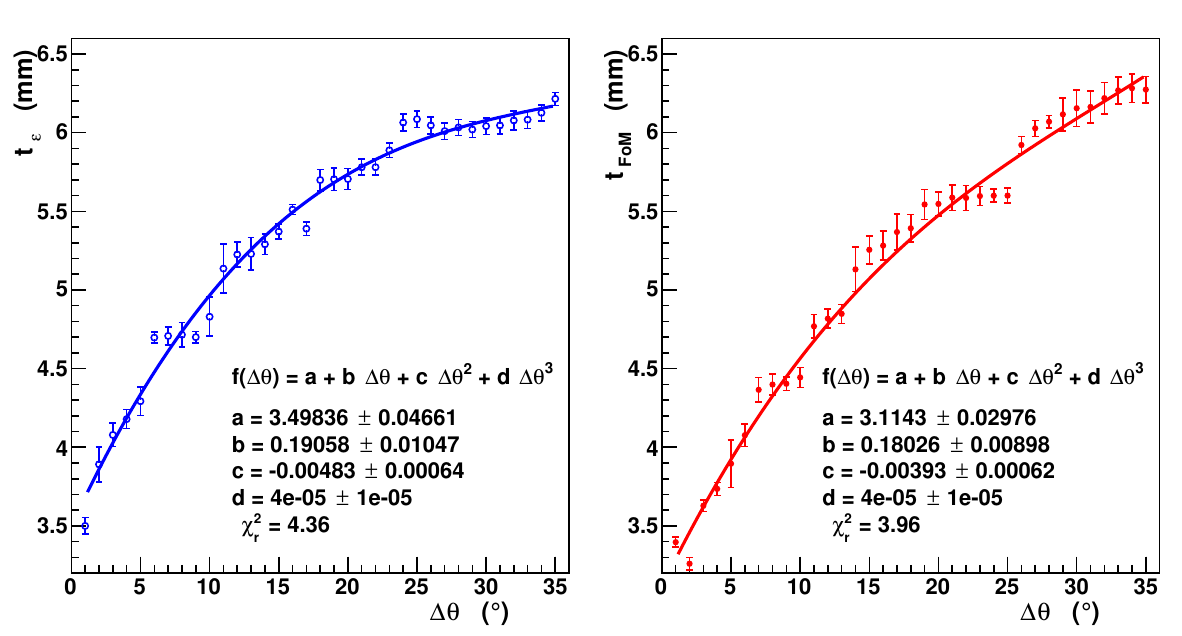}
\includegraphics[width=0.76\columnwidth]{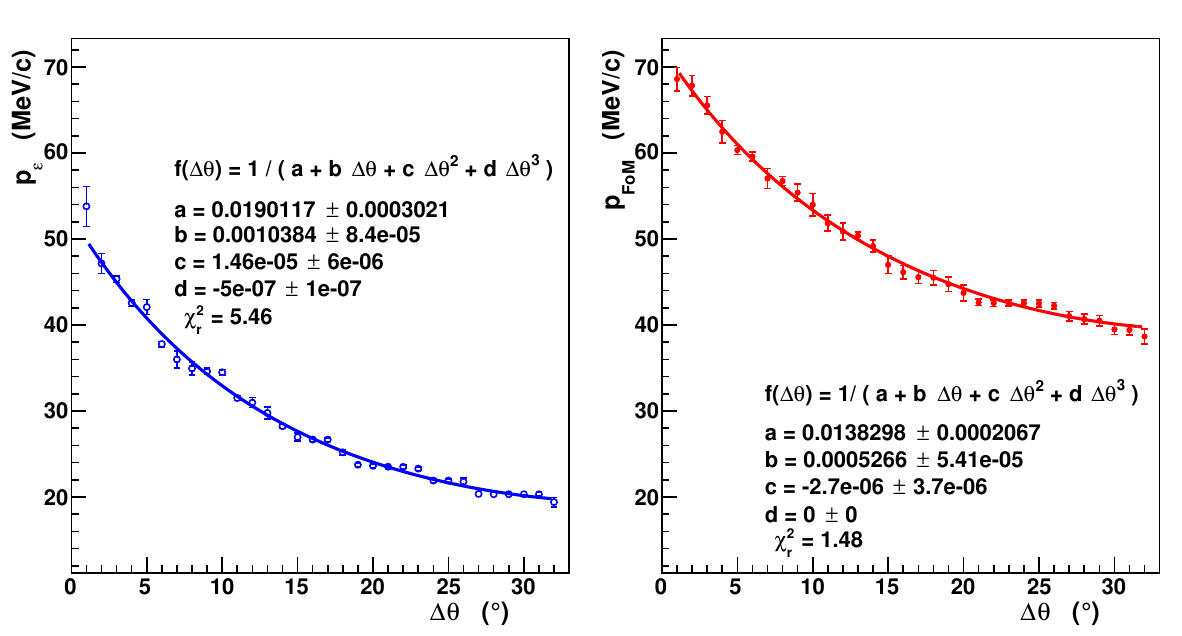}
\includegraphics[width=0.76\columnwidth]{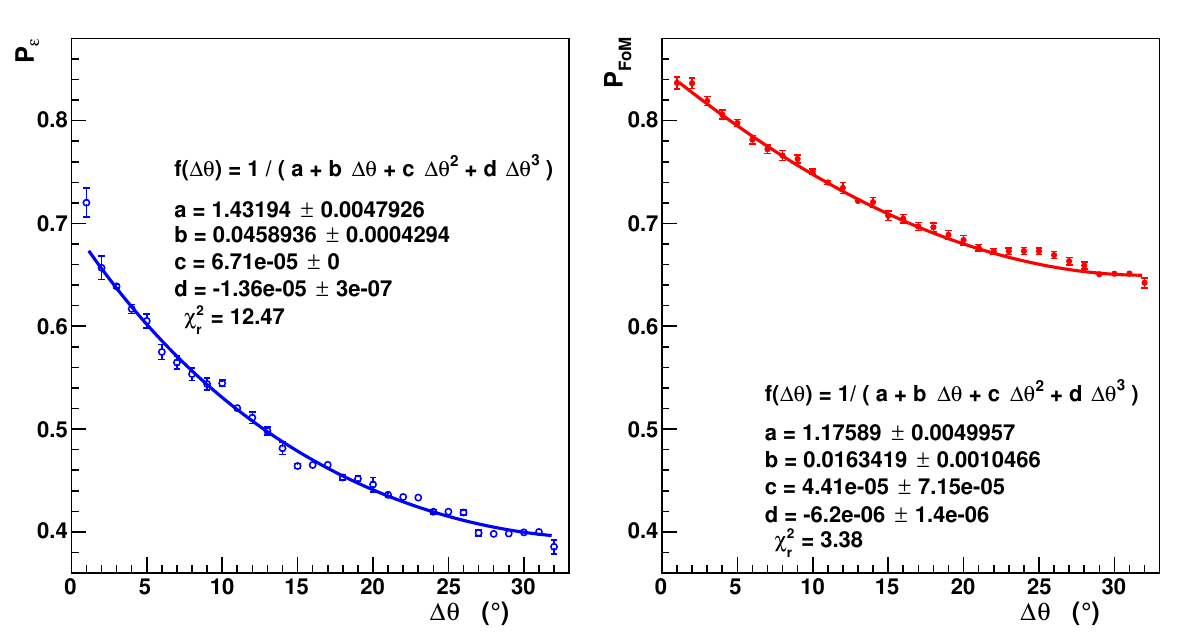}
\caption{Angular acceptance dependence of the optimum target thickness (top panel), the positron momentum at maximum (middle panel) and the positron polarization at maximum (bottom panel) of the unpolarized and polarized operational modes at 120~MeV.}
\label{thimompol120}
\end{figure}
\vfill\eject

% ----------------------------------------------------------------------------------------

\begin{table}[h!]
\centering
\begin{adjustbox}{max width=\textwidth}
\begin{tabular}{c||c|c|c|c|c|c||c|c|c|c|c|c}
 \multirow{2}{*}{$\Delta \theta$} & \multicolumn{6}{c||}{$\Delta p / p = \pm 5 $\%} & \multicolumn{6}{c}{$\Delta p / p = \pm 10 $\%} \\
& $t_{FoM}$ & $\delta t_{FoM}^-$ & $\delta t_{FoM}^+$ & $p_{FoM}$ & $\delta p_{FoM}^-$ & $\delta p_{FoM}^+$ & $t_{FoM}$ & $\delta t_{FoM}^-$ & $\delta t_{FoM}^+$ & $p_{FoM}$ & $\delta p_{FoM}^-$ & $\delta p_{FoM}^+$ \\
($^{\circ}$) & (mm) & (mm) & (mm) & (MeV/$c$) & (MeV/$c$) & (MeV/$c$) & (mm) & (mm) & (mm) & (MeV/$c$) & (MeV/$c$) & (MeV/$c$) \\ \hline\hline
          45 & 6.5 & 0.3 & 0.1 & 36.2 & 2.41 & 2.41 & 5.5 & 0.3 & 0.5 & 44.1 & 2.72 & 3.87 \\
          10 & 4.8 & 0.5 & 0.3 & 52.1 & 4.10 & 4.10 & 4.8 & 0.5 & 0.3 & 52.1 & 4.10 & 4.10 \\
\phantom{0}5 & 4.0 & 0.5 & 0.7 & 60.0 & 3.39 & 4.37 & 4.0 & 0.5 & 0.7 & 60.0 & 3.39 & 4.37 \\
\end{tabular}
\end{adjustbox}
\caption{Optimum target thickness $t_{FoM}$ and positron momentum at maximum $p_{FoM}$ of the polarized mode at 120~MeV, for different angular and momentum acceptances.}
\label{fom_max_120}
\end{table}

\subsection{Sensitivity to the collection system}
\label{sec:sense}

The positron collection system of the present optimization procedure is represented by momentum and angular acceptances which affect simulated data in different ways. The momentum acceptance is interpreted as the half-width of the momentum integrated positron yield, while the angular acceptance is an effective hard cut on the angular distribution of particles. The sensitivity to these parameters have been scanned in the $\Delta p/p$ range $\pm$1-15\% and the $\Delta \theta$ range $\pm$1-35$^{\circ}$. 

Consistently with observations of the previous sections, the $\Delta p/p$ dependence of the optimum thickness at fixed $\Delta \theta$ can be safely approximated by a constant. The value resulting from the fit of the $\Delta p/p$ distributions at fixed $\Delta \theta$ is represented on Fig.~\ref{thimompol120} (top panel) over the investigated angular domain and fitted to a 3rd-order polynomial. The expected larger statistical fluctuations at the smallest $\Delta \theta$'s are most likely responsible of a normalized $\chi^2_r > 1$. Globally, it is worth noticing that there exists a systematic difference between the optimum target thickness of the unpolarized and polarized modes, which tends to decrease with the angular acceptance. Similarly, the positron momentum and polarization at optimum do not depend on 
$\Delta p/p$ but are strongly sensitive to the angular acceptance. Following the same procedure, the $\Delta \theta$ dependencies are successfully fitted by the inverse of a 3rd-order polynomial on Figs.~\ref{thimompol120} (middle and bottom panels).

These features cannot be straightforwardly extended to the optimum efficiency and Figure-of-Merit in the sense that these quantities depend both on the momentum and angular acceptances. Nevertheless, it is obvious that both quantities are increasing functions of $\Delta p / p$ and $\Delta \theta$.

\section{Optimization of production at 1000~MeV}

\subsection{Unpolarized operational mode}

\begin{figure}[h!]
\centering
\includegraphics[width=0.99\columnwidth]{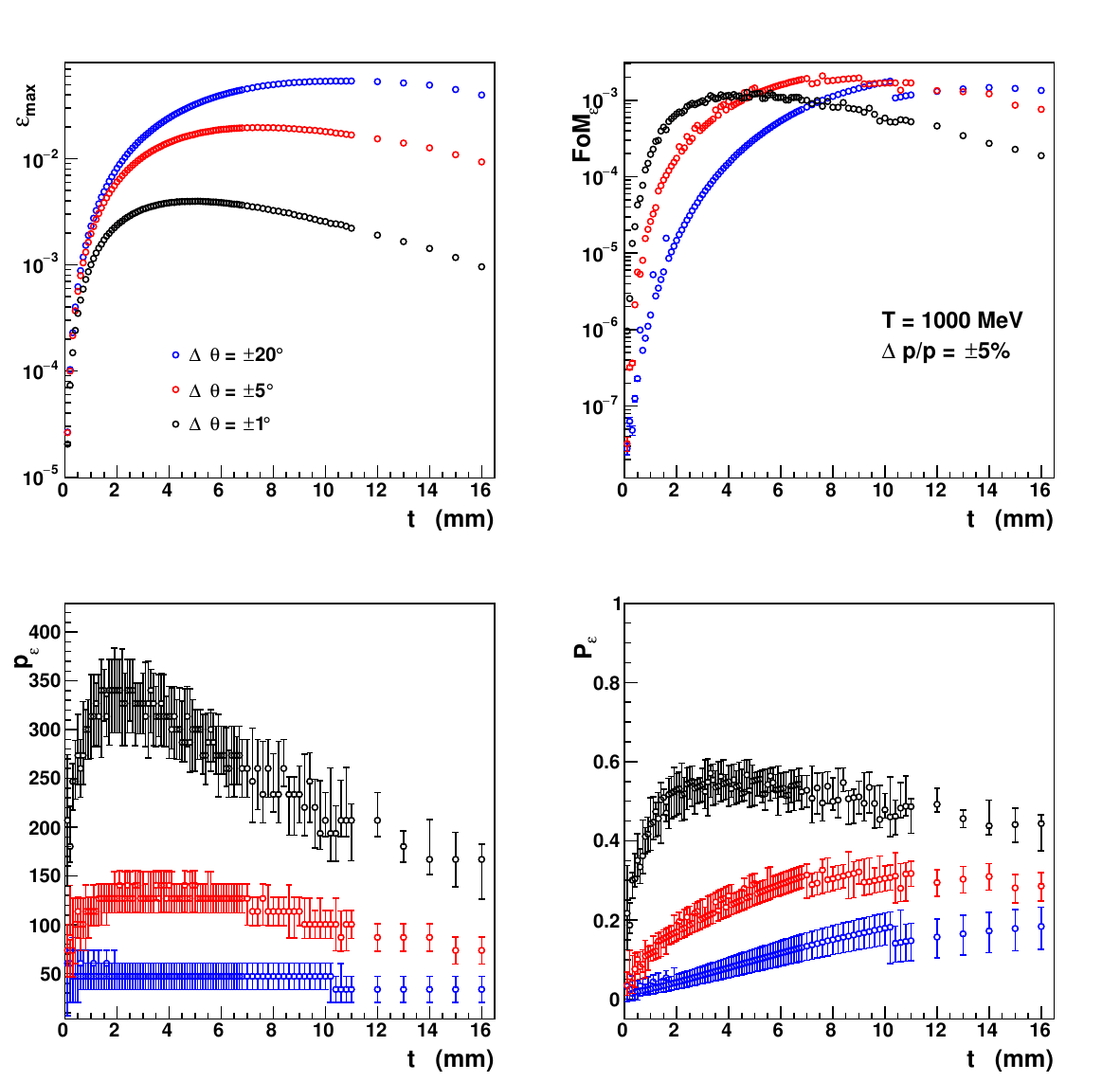}
\caption{Target thickness dependence of the characteristic  quantities of the unpolarized mode operation at 1000~MeV and a momentum acceptance of $\pm$5\% for different angular acceptances.}
\label{opt1000_unp}
\end{figure}

Similarly to the 120~MeV beam energy case, the maximum positron production efficiency is studied hereafter as function of the target thickness for an initial electron beam of 1000~MeV. A variable bin width corresponding to the positron momentum 
acceptance $\Delta p /p$=$\pm$5\% and different angular acceptances are considered. The results ($\varepsilon_{max}, FoM_{\varepsilon}, p_{\varepsilon}, P_{\varepsilon}$) are reported on Fig.~\ref{opt1000_unp} for target thicknesses up to 16~mm. The target thickness dependence of these characteristic 
parameters of the unpolarized operational mode is very similar to the 120~MeV case. The larger beam energy focuses the positron distribution which optimum angular domain is strongly reduced. For same acceptance conditions, larger production rates appears achievable with larger beam energies, however at the expense of a technologically more demanding collection system. 
As previously observed, $t_{\varepsilon}$ is barely sensitive to the momentum acceptance but strongly depends on the angular ones as reported in Tab.~\ref{eff_max_1000}.

% ----------------------------------------------------------------------------------------

\begin{table}[h!]
\centering
\begin{adjustbox}{max width=\textwidth}
\begin{tabular}{c||c|c|c|c|c|c||c|c|c|c|c|c}
 \multirow{2}{*}{$\Delta \theta$} & \multicolumn{6}{c||}{$\Delta p / p = \pm 5 $\%} & \multicolumn{6}{c}{$\Delta p / p = \pm 10 $\%} \\
& $t_{\varepsilon}$ & $\delta t_{\varepsilon}^-$ & $\delta t_{\varepsilon}^+$ & $p_{\varepsilon}$ & $\delta p_{\varepsilon}^-$ & $\delta p_{\varepsilon}^+$ & $t_{\varepsilon}$ & $\delta t_{\varepsilon}^-$ & $\delta t_{\varepsilon}^+$ & $p_{\varepsilon}$ & $\delta p_{\varepsilon}^-$ & $\delta p_{FoM}^+$ \\
($^{\circ}$) & (mm) & (mm) & (mm) & (MeV/$c$) & (MeV/$c$) & (MeV/$c$) & (mm) & (mm) & (mm) & (MeV/$c$) & (MeV/$c$) & (MeV/$c$) \\ \hline\hline
          20 & 11.0 & 0.4 & 1.0 & \phantom{1}33.8 & 13.4 & 13.4 & 10.8 & 0.8 & 1.2 & \phantom{1}33.8 & 13.7 & 13.7 \\
\phantom{0}5 & \phantom{1}7.4 & 0.4 & 0.6 & 113.7 & 14.5 & 14.5 & \phantom{1}7.4 & 0.2 & 0.6 & 113.7 & 17.5 & 17.5 \\
\phantom{0}1 & \phantom{1}5.1 & 0.3 & 0.2 & 300.2 & 30.6 & 20.1 & \phantom{1}5.1 & 0.5 & 0.2 & 300.2 & 40.1 & 40.1 \\
\end{tabular}
\end{adjustbox}
\caption{Optimum target thickness $t_{\varepsilon}$ and positron momentum at maximum $p_{\varepsilon}$ of the unpolarized mode at 1000~MeV, for different angular and momentum acceptances.}
\label{eff_max_1000}
\end{table}

\subsection{Polarized operational mode}

\begin{figure}[h!]
\centering
\includegraphics[width=0.99\columnwidth]{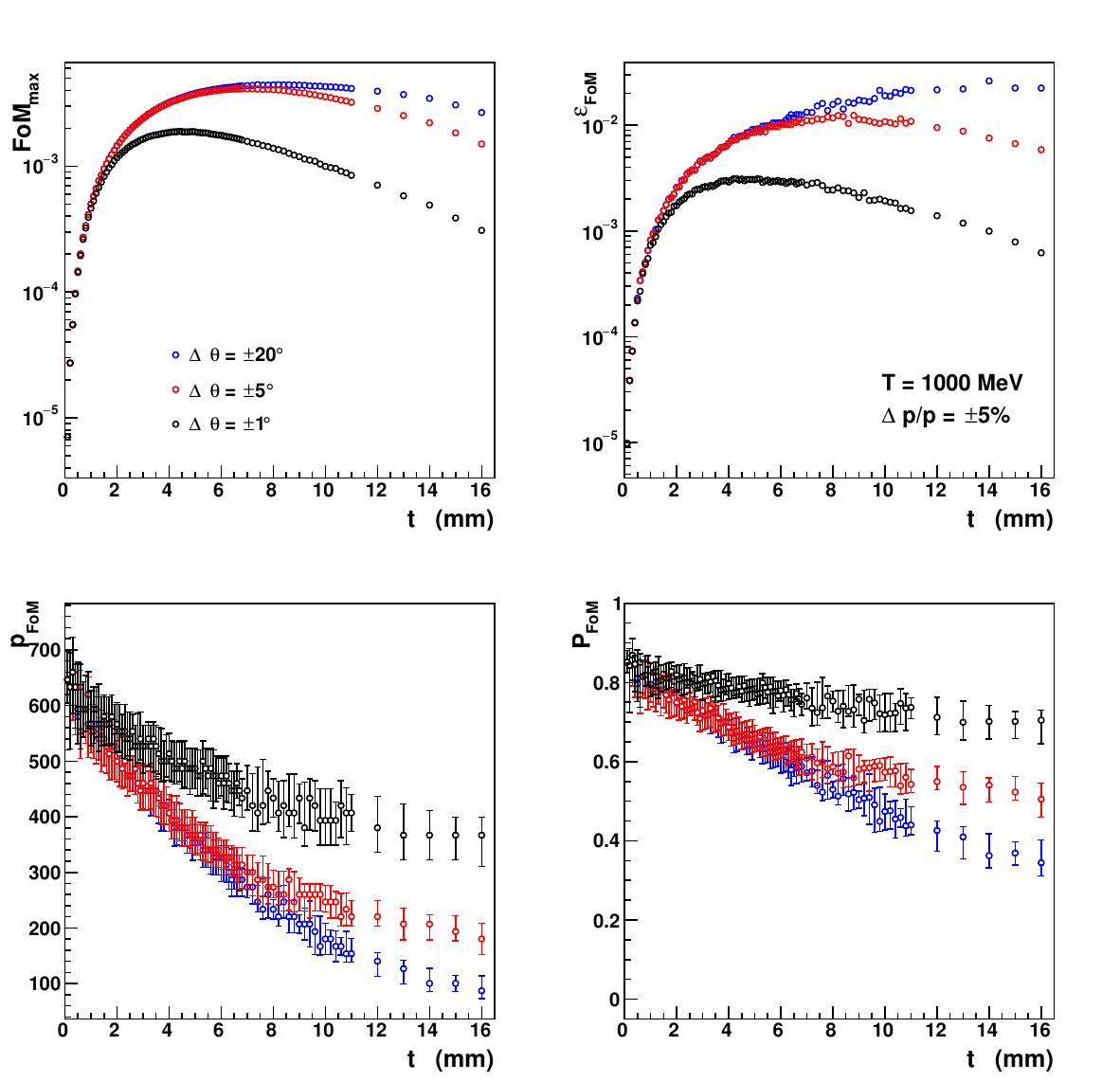}
\caption{Target thickness dependence of the characteristic  quantities of the polarized mode operation at 1000~MeV and a momentum acceptance of $\pm$5\% for different angular acceptances.}
\label{opt1000_pol}
\end{figure}

The dependence on the target thickness of the characteristic parameters of the polarized operational mode ($FoM_{max},\varepsilon_{FoM},p_{FoM}, P_{FoM}$) are investigated under the same conditions than the unpolarized mode (Fig.~\ref{opt1000_pol}). The comparison of the two operational modes does not bring new features than those observed at 120~MeV. There is globally no major differences between the low and high beam energies investigated here except that larger optimum $FoM$ are obtained, consistently with a larger production efficiency. The rule of thumb previously established~\cite{Dum11} is still valid, however with an enhanced sensitivity to the angular acceptance primarily due to the focusing effects of the initial electron beam.  
\begin{table}[h!]
\centering
\begin{adjustbox}{max width=\textwidth}
\begin{tabular}{c||c|c|c|c|c|c||c|c|c|c|c|c}
 \multirow{2}{*}{$\Delta \theta$} & \multicolumn{6}{c||}{$\Delta p / p = \pm 5 $\%} & \multicolumn{6}{c}{$\Delta p / p = \pm 10 $\%} \\
& $t_{FoM}$ & $\delta t_{FoM}^-$ & $\delta t_{FoM}^+$ & $p_{FoM}$ & $\delta p_{FoM}^-$ & $\delta p_{FoM}^+$ & $t_{FoM}$ & $\delta t_{FoM}^-$ & $\delta t_{FoM}^+$ & $p_{FoM}$ & $\delta p_{FoM}^-$ & $\delta p_{FoM}^+$ \\
($^{\circ}$) & (mm) & (mm) & (mm) & (MeV/$c$) & (MeV/$c$) & (MeV/$c$) & (mm) & (mm) & (mm) & (MeV/$c$) & (MeV/$c$) & (MeV/$c$) \\ \hline\hline
          20 & 7.4 & 1.6 & 4.6 & 246.9 & 18.2 & 29.4 & 8.2 & 2.2 & 2.8 & 233.6 & 35.4 & 35.4 \\
\phantom{0}5 & 7.2 & 1.0 & 0.8 & 300.2 & 42.7 & 30.6 & 7.2 & 0.8 & 0.4 & 300.2 & 40.1 & 40.1 \\
\phantom{0}1 & 4.4 & 0.6 & 0.9 & 500.0 & 36.5 & 36.5 & 4.9 & 1.0 & 0.5 & 486.7 & 55.5 & 55.5 \\
\end{tabular}
\end{adjustbox}
\caption{Optimum target thickness $t_{FoM}$ and positron momentum at maximum $p_{FoM}$ of the polarized mode at 1000~MeV, for different angular and momentum acceptances.}
\label{fom_max_1000}
\end{table}

\subsection{Sensitivity to the collection system}
\label{sec:1000}

Following the method described in Sec.~\ref{sec:sense} the sensitivity of the optimum target thickness - and therefore of the characteristics parameters of the positron source operational modes - to the acceptances of the collection system have been investigated in the $\Delta p/p$ range $\pm$1-15\% and the $\Delta \theta$ range $\pm$1-15$^{\circ}$. 

\begin{figure}[h!]
\centering
\includegraphics[width=0.76\columnwidth]{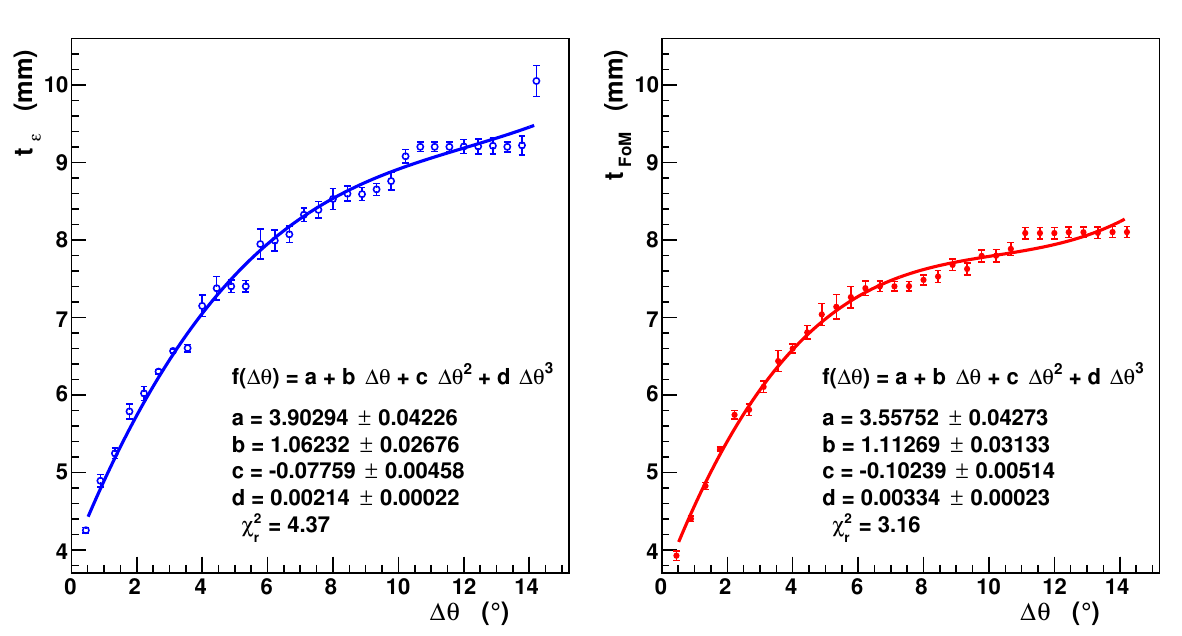}
\includegraphics[width=0.76\columnwidth]{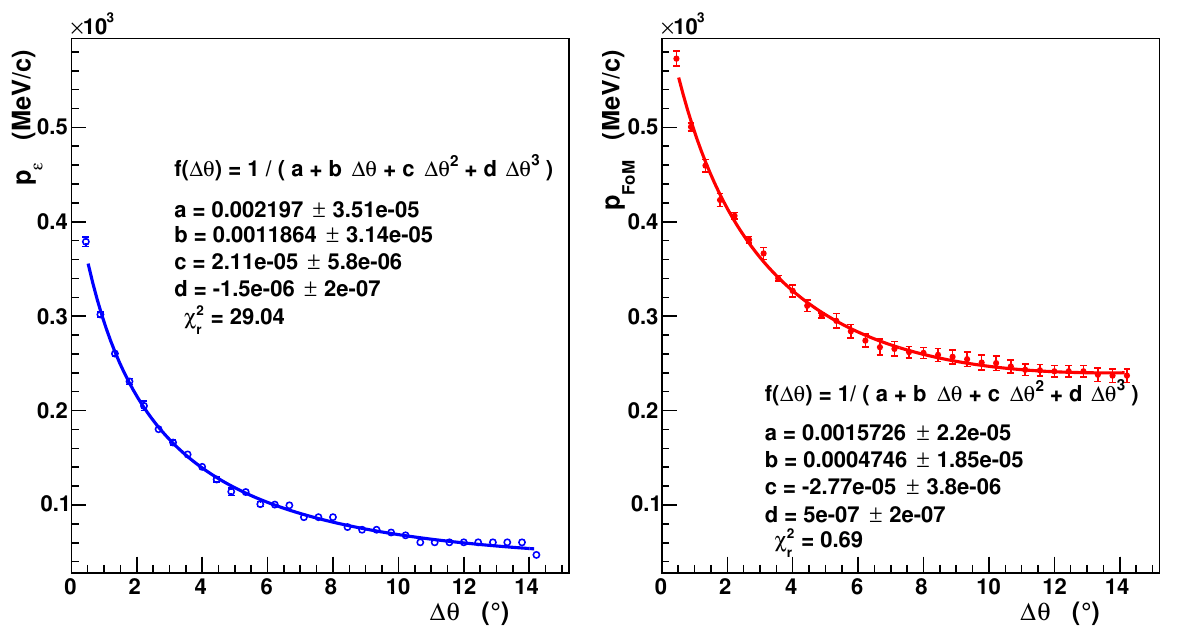}
\includegraphics[width=0.76\columnwidth]{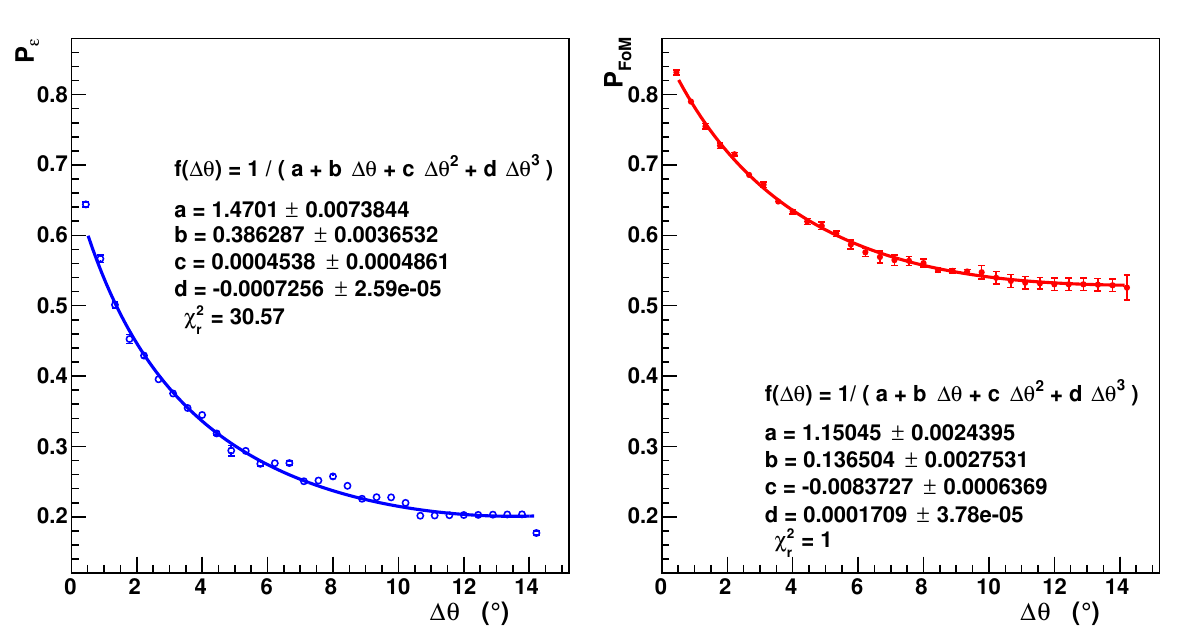}
\caption{Angular acceptance dependence of the optimum target thickness (top panel), the positron momentum at maximum (middle panel) and the positron polarization at maximum (bottom panel) of the unpolarized and polarized operational modes at 1000~MeV.}
\label{thimompol1000}
\end{figure}

The absence of correlations between the $\Delta p/p$ and $\Delta \theta$ dependencies allows to obtain the angular acceptance sensitivity of the optimum thickness (top panel) and of the positron momentum (middle panel) and polarization (bottom panel) at optimum (Fig.~\ref{thimompol1000}). While the $\Delta\theta$ dependencies of $t_{\varepsilon}$ and $t_{FoM}$ are similar, there exists a systematic difference between them which tends to increase with angular acceptance. 

\section{Electron beam energy dependence}

A precise comparison between the positron production at different beam energies is a difficult task since it strongly depends on the performances of the collection system which are not only linked to technological capabilities but also to the full set of elements required to match the positron beam with the acceptance of a further accelerator system. In that respect, representing the collection system by simple momentum and angular acceptances should not be considered more than a fair approximation. 

% ----------------------------------------------------------------------------------------

\newpage

\null\vfill
\vfill\eject

\null\vfill
\vfill\eject

\null\vfill
\vfil\eject

% ----------------------------------------------------------------------------------------

Two collection systems have been so far evaluated~\cite{Hab20}: the Quarter Wave Transformer (QWT) and the Adiabatic Matching Device (AMD). Both systems are characterized by a maximum ($B_1$) and a minimum ($B_2$) solenoidal field, and a high field region of length $L_1$. The QWT acts as a momentum filtering by favoring the collection of particles with momentum
\begin{equation}
p_0 = \frac{e B_1 L_1}{\pi}
\end{equation}
within a momentum acceptance
\begin{equation}
\frac{\Delta p}{p_0} = \frac{4}{\pi} \, \frac{B_2}{B_1} \,.
\end{equation}
The AMD does not filter the momentum, that is the momentum selection required for a PEPPo-like positron source is made further down the beam line by the selection chicane~\cite{Hab22}. The radius $r_0^{QWT}$ of the positron beam spot accepted by the QWT is linked to the radius $R$ of the low field section according to the relation
\begin{equation}
r_0^{QWT} = \frac{B_2}{B_1} \, R \label{eqT1}
\end{equation}
with a maximum accepted transverse momentum
\begin{equation}
p_t^{QWT} = \frac{e B_1 R}{2} \, . \label{eqT2}
\end{equation}
The corresponding angular acceptance can be expressed as 
\begin{equation}
\Delta \theta^{QWT} = \frac{\pi}{2} \, \frac{R}{L_1}
\end{equation}
which only depends on the length of the high field region. Comparing the transverse acceptances of the QWT (Eqs.~(\ref{eqT1})-(\ref{eqT2})) and the AMD leads to the relations
\begin{eqnarray}
r_0^{AMD} & = & \sqrt{\frac{B_1}{B_2}} \, r_0^{QWT} \\
p_t^{AMD} & = & \sqrt{\frac{B_2}{B_1}} \, p_t^{QWT} \, ,
\end{eqnarray}
indicating a larger transverse size acceptance but a smaller transverse momentum one and consequently a smaller angular acceptance for a given momentum of interest. 

Assuming the same technology ({\it i.e.} same $B_1$ and $B_2$) for positron collection at different beam energies, the QWT selection of larger momenta requires a longer high field region such that
\begin{equation}
L_1^{1000} = \frac{p_0^{1000}}{p_0^{120}} \, L_1^{120} \, ,
\end{equation}
which in turns implies a smaller angular acceptance
\begin{equation}
\Delta \theta^{1000} = \frac{p_0^{120}}{p_0^{1000}} \, \Delta \theta^{120} \, .
\end{equation}
Taking into account the correlation between the optimum momentum and the angular acceptance (Secs.~\ref{sec:sense}-\ref{sec:1000}), the optimum performance characteristics at 1000~MeV and 120~MeV are compared in Tab.~\ref{EneComp}. The benefit of larger electron beam energies is striking, particularly when the angular acceptance increases. This suggests that increasing the beam energy is one path to follow to reach high positron beam intensities. However, this path remains constrained by the total sustainable beam power. Another somehow equivalent path is to enlarge the angular acceptance of the collection system, that is to reduce the high field region lentgh $L_1$ and compensate with higher $B_1$ field to select similar positron momentum. In that respect, technological limitations on $B_1$ which affect more rapidly the polarized mode, favor lower beam energies to allow optimum operation with the same collection device and beam line for both the unpolarized and polarized modes. 

\begin{table}[t!]
\centering
\begin{adjustbox}{max width=\textwidth}
\begin{tabular}{c||c|c|c|c|c|c||c|c|c|c|c|c}
$T$ & $\Delta \theta$ & $t_{\varepsilon}$ & $p_{\varepsilon}$ & $P_{\varepsilon}$ & $\varepsilon_{max}$ & $FoM_{\varepsilon}$ & $\Delta \theta$ & $t_{FoM}$ & $p_{FoM}$ & $P_{FoM}$ & $\varepsilon_{FoM}$ & $FoM_{max}$ \\
(MeV) & ($^{\circ}$) & (mm) & (MeV/$c$) & (\%) & ($\times 10^{-3}$) & ($\times 10^{-3}$) & ($^{\circ}$) & (mm) & (MeV/$c$) & (\%) & ($\times 10^{-3}$) & ($\times 10^{-3}$) \\ \hline\hline
1000 & \phantom{1}1.0 & 4.9 & 293.9 & 53.9 & 3.97 & 1.20 & \phantom{1}1.0 & 4.6 & 495.0 & 78.2 & 2.99 & 1.87 \\ 
\phantom{1}120 & \phantom{1}7.9 & 4.7 & 35.82 & 55.8 & 2.44 & 0.72 & \phantom{1}9.1 & 4.5 & 54.34 & 75.5 & 2.22 & 1.26 \\ \hline\hline 
1000 & \phantom{1}3.0 & 6.4 & 169.3 & 38.3 & 12.5 & 1.76 & \phantom{1}3.0 & 6.1 & 362.2 & 67.1 & 7.88 & 3.61 \\ 
\phantom{1}120 & 17.5 & 5.6 & 25.63 & 45.8 & 5.47 & 1.22 & 29.3 & 6.0 & 37.14 & 65.1 & 5.52 & 2.33 \\
\end{tabular}
\end{adjustbox}
\caption{Comparison of optimum performances of a PEPPo-like positron source operating at different beam energies, considering the technological constraints of the collection system and a momentum acceptance of $\pm$5\%.}
\label{EneComp}
\end{table}

% ----------------------------------------------------------------------------------------

\section{Conclusion}

The production of polarized positrons on a tungsten target has been investigated for two incident electron beam energies. An optimization method was proposed and developed which allows to obtain an optimum efficiency or Figure-of-Merit depending on the momentum and angular acceptances of a collection system. The optimal operating conditions - in terms of the target thickness and of the positron momentum and polarization at optimum - have been found insensitive to the momentum acceptance but strongly depending on the angular one. Together with the technological constraints on the maximum achievable magnetic field of the collection system, the angular acceptance appears the most important parameter in order to reach high positron beam intensities in a high duty cycle PEPPo-like positron source to operate at CEBAF.

% ----------------------------------------------------------------------------------------

\section*{Acknowledgements}

This research work is part of a project that has received funding from the European Union's Horizon 2020 research and innovation program under agreement STRONG - 2020 - No~824093. 
%It was supported by the
%U.S. Department of Energy, Office of Science, Office of
%Nuclear Physics, contract DE-AC05-06OR23177 under
%which Jefferson Science Associates, LLC operates JLab.

% ----------------------------------------------------------------------------------------

% ----------------------------------------------------------------------------------------

\end{document}